\documentclass[aip,rsi,superscriptaddress,amsmath,amssymb,reprint,longbibliography]{revtex4-2}

\usepackage{graphicx}
\usepackage{placeins}
\usepackage{lettrine}
\usepackage{adjustbox}
\usepackage{multirow}
\usepackage{xcolor}
\usepackage{bm}
\usepackage[colorlinks=true,
            linkcolor=blue,
            citecolor=blue,
            urlcolor=blue]{hyperref}

\begin{document}
\title{Evidence for Vortex Rings with Multiquantum Circulation in He~II}


\author{Yiming Xing}\thanks{These authors contributed equally to this work.}
\affiliation{National High Magnetic Field Laboratory, Tallahassee, FL 32310, USA}
\affiliation{Mechanical and Aerospace Engineering Department, FAMU-FSU College of Engineering, Florida State University, Tallahassee, FL 32310, USA}

\author{Yousef Alihosseini}\thanks{These authors contributed equally to this work.}
\affiliation{National High Magnetic Field Laboratory, Tallahassee, FL 32310, USA}
\affiliation{Mechanical and Aerospace Engineering Department, FAMU-FSU College of Engineering, Florida State University, Tallahassee, FL 32310, USA}

\author{Sosuke Inui}\thanks{These authors contributed equally to this work.}
\affiliation{National High Magnetic Field Laboratory, Tallahassee, FL 32310, USA}
\affiliation{Mechanical and Aerospace Engineering Department, FAMU-FSU College of Engineering, Florida State University, Tallahassee, FL 32310, USA}

\author{Wei Guo}
\email[Corresponding: ]{wguo@magnet.fsu.edu}
\affiliation{National High Magnetic Field Laboratory, Tallahassee, FL 32310, USA}
\affiliation{Mechanical and Aerospace Engineering Department, FAMU-FSU College of Engineering, Florida State University, Tallahassee, FL 32310, USA}

\date{\today}

\begin{abstract}
Quantized vortex dynamics in superfluid $^4$He (He~II) are widely regarded as well established: circulation is quantized in units of $\kappa=h/m_4$, vortices carrying more than one quantum are expected to split into singly quantized filaments, and vortex rings shrink while accelerating due to dissipation from thermal-quasiparticle scattering. Using particle tracking velocimetry with frozen deuterium tracers, we uncover rare vortex-bound particle events that disrupt this canonical picture. In a class of events exhibiting the acceleration characteristic of shrinkage driven vortex ring motion, the measured kinematics cannot be reconciled with a singly quantized ring. Instead, they require an effective circulation $n\kappa$ with $n>1$, directly challenging the standard expectation that multiquantum vortices are short lived. A more prosaic possibility is that the inferred $n\kappa$ arises from a bundle of closely spaced singly quantized rings, which could generate similar large-scale motion. However, this scenario is disfavored by vortex-filament simulations that show rapid bundle dispersion. Furthermore, the persistence of particle trapping at the observed high speeds suggests a much deeper core trapping potential, consistent only with a truly multiquantum core. Together, these results point to anomalously long-lived multiquantum rings, a striking puzzle that calls for dedicated scrutiny beyond the prevailing paradigm.
\end{abstract}
\maketitle

Quantized vortices govern dynamics and dissipation in a wide range of systems, from atomic Bose–Einstein condensates (BECs) to type-II superconductors, from rotating superfluid neutron stars to topological defects in cosmological settings~\cite{Tsubota-2013-book, RevModPhys.66.1125, Greenstein-1970-Nature,pines1985superfluidity,zurek1985cosmological}. Despite this diversity, the same vortex-level mechanisms recur: circulation is quantized, vortices drift and exchange topology via reconnections, and finite-temperature dissipation is mediated by thermal quasiparticles scattering from vortex cores~\cite{tilley2019superfluidity}. A laboratory platform that exposes these mechanisms through quantitative measurements of vortex trajectories can therefore deliver insight with broad reach.

Superfluid $^4$He (He II) has long been regarded as such a platform. In Landau’s two fluid framework~\cite{Landau1941PR}, an inviscid superfluid component coexists with a viscous normal component composed of thermal quasiparticles. Any superfluid vorticity is confined to quantized vortex lines with a core radius $a_0\simeq 1$~\AA\ and a fixed circulation $\kappa=h/m_4$, where $h$ is Planck’s constant and $m_4$ is the mass of a $^4$He atom~\cite{tilley2019superfluidity,Landau1941PR}. This separation of scales, i.e., angstrom core and macroscopic flow field, makes vortex structures in He II exceptionally well defined and amenable to controlled modeling~\cite{vinen2002quantum}. At finite temperature, quasiparticle scattering off vortex cores produces mutual friction \cite{vinen1957mutual,Barenghi1983}, whose established formulation allows quantitative prediction of the vortex motion \cite{Schwartz1985}. In parallel, tracer particle visualization has opened a direct window onto vortex dynamics in real space, revealing key features of both individual vortices and vortex tangles~\cite{Bewley-2006-Nature, Guo-2014-PNAS, Mantia-PRB-2014, Mastracci-2019-PRF, Fonda-2019-PNAS, Tang-2023-NatCommun, Peretti2023SciAdv, Minowa2025NatPhys}. Together, these advances have created an unusually coherent picture in which quantitative measurements and established theory often align, reinforcing the common view that the essential vortex physics in He II is largely settled for most experimental conditions.

In particular, it is widely accepted that stable circulation in He~II must be singly quantized. This expectation follows directly from an energetic scaling argument. For instance, for a singly quantized vortex ring of radius $R$, the kinetic energy of the associated flow field is $E\simeq \frac{\rho_{\rm s}\kappa^2R}{2}\left[\ln(8R/a_0)-\frac{3}{2}\right]$~\cite{Donnelly-1991-B}. A ring of the same radius but carrying $n$ quanta of circulation therefore has an energy approximately given by $n^2E$. On the other hand, $n$ well separated singly quantized rings have a total energy of $nE$. The resulting excess energy, $(n^2-n)E>0$ for $n>1$, makes multiquantum circulation energetically unfavorable and leads to the standard expectation that it should rapidly split into singly quantized vortices under generic perturbations.



\begin{figure*}[t!]
\centering
\includegraphics[width=1\linewidth]{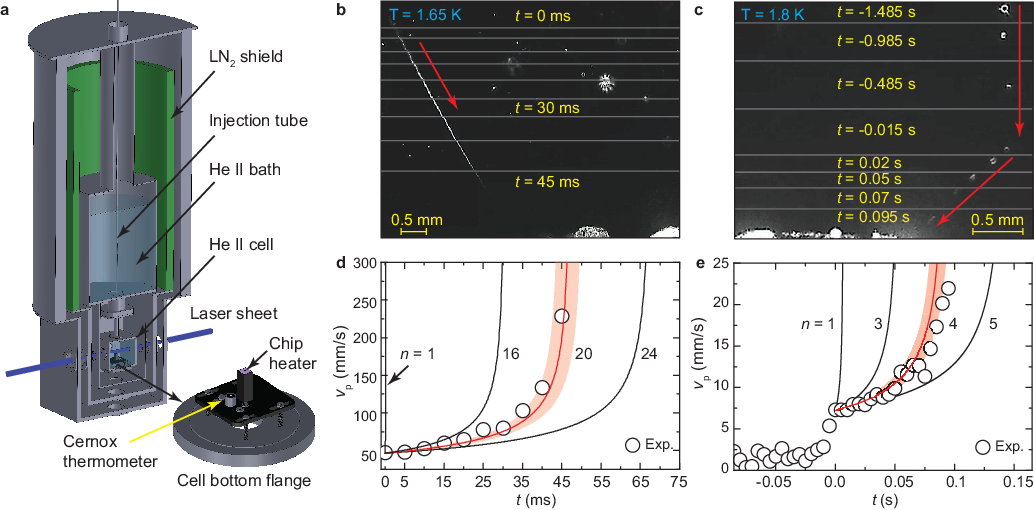}
\caption{\textbf{Vortex-bound particle events suggesting multiquantum vortex-ring kinematics.} \textbf{a} Schematic of the experimental setup. \textbf{b,c} Superimposed image sequences for two representative events recorded at heater-surface voltages $V_\mathrm{H}=0$~V and $+10$~V, respectively; red arrows indicate the direction of motion. \textbf{d,e} Measured particle speed $v_p(t)$ along the trajectories in \textbf{b} and \textbf{c}. Solid curves show the predicted ring propagation speed $v_{\parallel,n}(t)$ for various circulations $n\kappa$, with each curve initialized by matching the starting speed $v_p(0)$. Shaded bands represent the resulting spread in $v_{\parallel,n}(t)$ due to the uncertainty in $v_p(0)$.}
\label{Fig1}
\end{figure*}


Here we report rare vortex-bound particle events observed in our particle-tracking velocimetry (PTV) experiments using micron-sized frozen deuterium (D$_2$) tracers that force a re-examination of this prevailing picture. The particle trajectories display the rapid acceleration expected for a shrinking vortex ring \cite{Donnelly-1991-B}. However, quantitative analysis of the measured kinematics rules out a singly quantized ring and instead requires an effective circulation $n\kappa$, with $n$ in the range 3 to 20. We further disfavor an interpretation in terms of a compact bundle of singly quantized rings, since such bundles disperse rapidly under mutual friction and are incompatible with sustained particle trapping at the observed high speeds. In what follows, we present our observations and discuss the analysis that leads to this intriguing conclusion.

The experimental setup is shown in Fig.~\ref{Fig1}a. A D$_2$/$^4$He gas mixture is injected through a stainless steel tube into an experimental cell connected to the He II bath~\cite{Mastracci-2018-RSI,Tang-2021-PNAS}; the bath temperature is controlled via vapor-pressure regulation and monitored by a Cernox sensor. Following injection, the D$_2$ solidifies into ice particles with a mean radius $a_p\simeq1.1$~$\mu$m~\cite{Tang-2023-NatCommun}, inferred from their settling velocities in quiescent He II (see Methods). As particles approach vortex cores, they can become trapped by the Bernoulli pressure associated with the circulating superfluid flow~\cite{Donnelly-1991-B}, thereby directly revealing vortex structures. In some runs, a chip-resistor heater (area: $1.6\times0.5$~mm$^2$) in the cell is pulsed on for a few milliseconds to promote vortex generation. We illuminate the tracer particles with a thin laser sheet ($\sim$150~$\mu$m thickness) and record their motion at 200~Hz with a camera placed perpendicular to the sheet. When a vortex event passes through the laser plane, the coordinates of the trapped particles are extracted using a feature-point tracking algorithm~\cite{Sbalzarini-2005-JSB}, enabling quantitative kinematic analysis.

In our experiments, the tracers most often outline relatively smooth vortex filaments that drift slowly through the laser sheet and occasionally undergo reconnections as they cross~\cite{Stasiak-2025-PNAS}. We also observe regular vortex rings decorated by a few tracers propagating through otherwise quiescent He~II~\cite{Tang-2023-NatCommun}. Beyond these understood structures, a distinct class of rare events were captured over a broad temperature range, typically involving a single tracer moving through He~II at high speed and, most strikingly, \emph{accelerating} with time (see Supplementary Movie~1 for a collection of example events).

Two representative trajectories are shown in Fig.~\ref{Fig1}b and \ref{Fig1}c (see also Supplementary Movies~2 and 3), and the corresponding particle speed $v_p(t)$ measured along the trajectory direction is shown in Fig.~\ref{Fig1}d and \ref{Fig1}e. The rapid growth of $v_p(t)$ strongly suggests vortex ring kinematics. In a static He~II background, a singly quantized vortex ring of radius $R$ propagates along its symmetry axis with speed $v_{\parallel}(R)\simeq\frac{\kappa}{4\pi R}\left[\ln\left(\frac{8R}{a_0}\right)-\frac{1}{2}\right]$. As $R$ decreases due to mutual friction dissipation, the ring naturally speeds up. In Schwarz’ model, the shrinkage is governed by $dR/dt\simeq-\alpha(T)v_{\parallel}(R)$~\cite{Schwartz1985}, where $\alpha(T)$ is a temperature-dependent mutual friction coefficient~\cite{vinen1957mutual,Barenghi1983}. More refined models that include mutual friction feedback on the normal fluid are available~\cite{Kivotides-2000-Science, Yui-2020-PRL, Galantucci-2020-EPJ}, but they predict only mild changes in the ring kinematics~\cite{Tang-2023-NatCommun}. In what follows, we adopt the Schwarz model, as it provides compact easy-to-follow expressions.

We focus on events whose in-plane trajectory length is far larger than the laser-sheet thickness, so that the ring’s propagation direction lies nearly within the sheet and the observed speed closely tracks $v_{\parallel}$. By analyzing the particle velocity over the first few frames, we can determine the initial speed $v_p(0)$ (see Methods). Using the expression for $v_{\parallel}(R)$, we then infer the corresponding initial radius $R(0)$ for each event. Starting from this $R(0)$, we evolve the ring by solving the shrinkage-rate equation, obtaining $R(t)$ and the associated $v_{\parallel}(R(t))$ self-consistently. The resulting velocity curves $v_{\parallel}(t)$ for singly quantized rings are shown in Fig.~\ref{Fig1}. Surprisingly, they are in stark disagreement with the measured $v_p(t)$. The root of the failure is the large initial speed $v_p(0)$ in these events. For instance, for the event shown in Fig.~\ref{Fig1}(b) with $v_p(0)\simeq 5$~cm/s, the inferred $R(0)\approx 1.8~\mu$m. Such a small ring would be rapidly damped by mutual friction and shrink away on a timescale far too short to account for the observation.

This mismatch motivated us to consider alternative explanations for the observed accelerating motion. In particular, many of the events occur in runs following a heat pulse from the heater, raising the concern that the tracers are simply charged and are being pulled toward the heater surface by a residual electric field. To test this, we deliberately grounded the heater or applied constant voltages to its surface; neither the event rate nor the trajectory pattern showed any systematic dependence on voltage polarity or magnitude. In the particularly striking event shown in Fig.~\ref{Fig1}(c), a particle first settles slowly in quiescent He~II and then is abruptly entrained into an accelerating, ring-like motion toward the heater surface, which is held at $+10$~V. If one nevertheless interprets the motion as electric-field-driven drift, the required charge can be estimated by balancing the electric force $F_E=QE$ with the Stokes drag $F_d=6\pi\mu_n a_p v_p$~\cite{Landau-book}, which yields $Q\simeq-8\times 10^2 e$ for the known local field $E$ (see Methods). Such an abrupt acquisition of hundreds of electron charge $e$ is unrealistic, especially because electrons in He~II form individual bubbles that do not readily coalesce into a multicharge packet that could be captured all at once~\cite{Guo2007_JLTP}. Consistently, separate tests using a uniform field generated by two large plate electrodes showed no measurable drift of the injected D$_2$ particles, confirming that they are effectively neutral.

With electrostatic drift ruled out, we now consider a more radical possibility: vortex rings carrying multiple quanta of circulation $n\kappa$. The corresponding self-induced speed is $v_{\parallel,n}(R)\simeq \frac{n\kappa}{4\pi R}\!\left[\ln\!\left(\frac{8R}{na_0}\right)-\frac{1}{2}\right]$. Here we take the effective core size to be $na_0$, based on the criterion of Glaberson \emph{et al.} that the core emerges when the azimuthal superflow reaches the Landau critical velocity~\cite{Glaberson1968,Barenghi1983}. This change provides only a minor logarithmic correction. Fitting the initial particle speed $v_p(0)$ then yields an initial radius $R(0)$ that is about $n$ times larger than in the singly quantized case. The shrinkage-rate equation now reads $dR/dt\simeq-\alpha(T,n)v_{\parallel,n}(R)$. To obtain the effective mutual-friction coefficient $\alpha(T,n)$ for an $n$-quanta core, we leverage the model of Samuels and Donnelly for the microscopic mutual-friction coefficients~\cite{Samuels-PRL-1990}, which provides a clear framework for how $\alpha$ scales with circulation~\cite{Barenghi1983,Donnelly-1991-B}. As shown in Methods, $\alpha(T,n)$ decreases with $n$ and, for large $n$, scales to leading order as $1/n$. As a result, the shrinkage rate becomes largely insensitive to $n$. The larger initial radius therefore allows the ring to persist for much longer times. In Fig.~\ref{Fig1}d and \ref{Fig1}e, we plot the predicted $v_{\parallel,n}(t)$ for various $n$ values. Strikingly, the measured $v_p(t)$ for the two cases shown is well reproduced by $n=20$ and $n=4$, respectively. Likewise, the other events can also be matched well with $n>1$. In these analyses, we neglect effects due to the single trapped D$_2$ particle on the vortex core (e.g., Stokes drag and weight), since they are only a few percent of the mutual friction experienced by the rings (see Methods).

\begin{figure}[t!]
\centering
\includegraphics[width=1\linewidth]{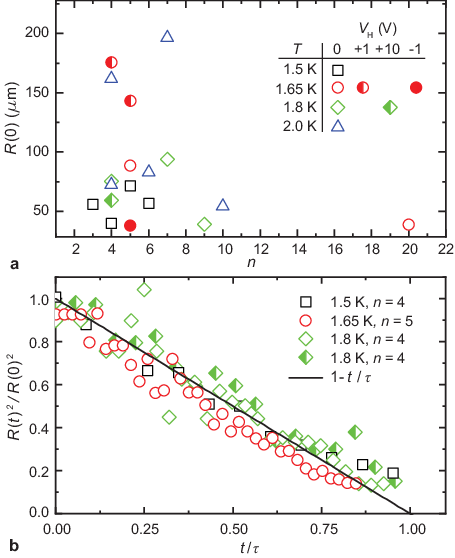}
\caption{\textbf{Deduced parameters of multiquantum vortex-ring events.} \textbf{a} Best-matched circulation number $n$ and initial radius $R(0)$ for all events collected at different temperatures and heater-surface voltages $V_\mathrm{H}$. \textbf{b} Normalized radius evolution, $R^2(t)/R^2(0)$, plotted versus normalized time $t/\tau$ for representative events. Here $R(t)$ is inferred from the particle speed $v_p(t)$, and the time scale $\tau$ is defined in the text.}
\label{Fig2}
\end{figure}

In Fig.~\ref{Fig2}(a), we compile all multiquantum ring events collected to date and show the best-matched circulation number $n$ together with the inferred initial radius $R(0)$. Most events cluster at $n<10$ with $R(0)\simeq 50$--$200~\mu$m. By treating the logarithmic factor in $v_{\parallel,n}(R)$ as a constant, one can integrate the shrinkage-rate equation and obtain a compact universal scaling~\cite{Donnelly-1991-B}: $R^2(t)/R^2(0)\simeq 1-t/\tau$, where $\tau \equiv R^{2}(0)/\frac{\alpha(T,n)\,n\kappa}{2\pi}\!\left[\ln\!\left(\frac{8R(0)}{na_0}\right)-\frac{1}{2}\right]$ (see Methods). In Fig.~\ref{Fig2}(b), we show $R^2(t)/R^2(0)$ for representative events, where $R(t)$ is inferred from the measured $v_p(t)$ using the $v_{\parallel,n}(R)$ expression. The data taken under various conditions all collapse onto this universal curve, strongly suggesting that these events are governed by vortex-ring dynamics.

\begin{figure}[t!]
\centering
\includegraphics[width=1\linewidth]{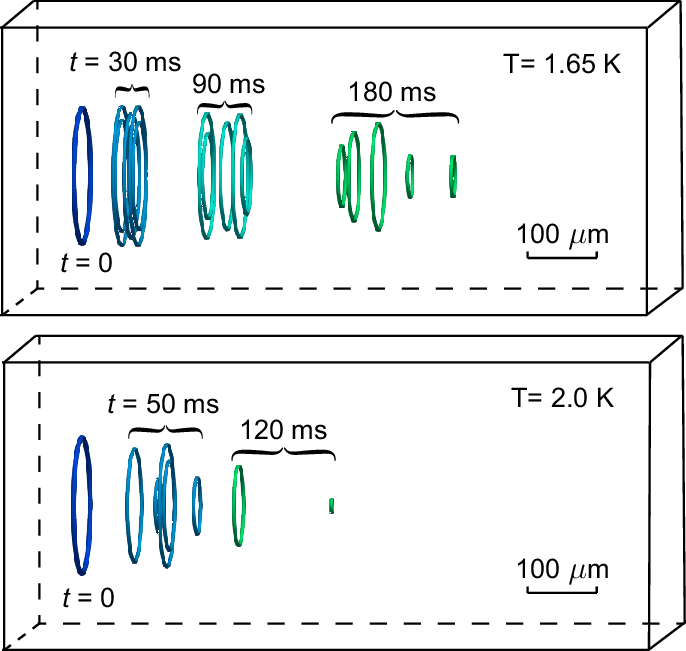}
\caption{\textbf{Schwarz model simulations showing rapid dispersion of singly quantized vortex-ring bundles.} For both cases at $T=1.65$~K and $2.0$~K, the bundle is initialized with five vortex rings of mean radius $R(0)=100$~$\mu$m and an inter-ring spacing of 2~$\mu$m. The bundles at later times are rendered in different colors for better visibility.}
\label{Fig3}
\end{figure}

One may argue that our observations do not require a truly multiquantum core, but could instead be explained by a bundle of closely spaced singly quantized rings, since such a bundle can generate a flow field resembling a multiquantum ring on length scales larger than the inter-ring spacing. However, our simulations, using the Schwarz model (see Methods), indicate that such compact bundles disperse under mutual friction over distances far shorter than the observed trajectories (see Fig.~\ref{Fig3} and Supplementary Movies 4 and 5). It is worth noting that Galantucci \emph{et al.}, using the refined FOUCAULT model, revealed that mutual-friction feedback can drive eddies in the normal fluid that help sustain a coherent bundle profile~\cite{galantucci-2023-PRF}. For instance, they showed that a bundle of 169 rings with $R(0)=120$~$\mu$m can persist out to a distance of $\sim10R(0)$, which is nevertheless still shorter than the typical trajectory lengths we observe. Moreover, their preliminary studies on few-ring bundles indicate even more rapid dispersion. Together, these results make simple coherent-ring bundles an unlikely explanation.

\begin{figure}[t!]
\centering
\includegraphics[width=1\linewidth]{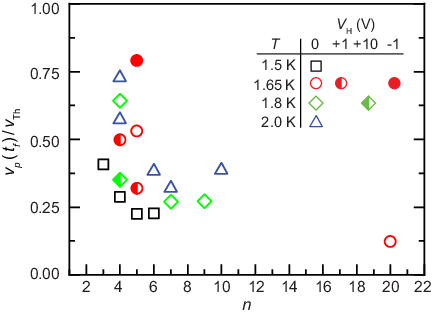}
\caption{\textbf{Ratio of the observed final particle speed $v_p(t_f)$ to the drag-limited detrapping threshold velocity $v_{th}=\rho_s(n\kappa)^2/18\pi^2 a_p\mu_n$}.}
\label{Fig4}
\end{figure}

On the other hand, the multiquantum-ring picture can naturally resolve another puzzle: how the D$_2$ particles remain trapped despite their large speeds. For a particle bound to a singly quantized vortex core, the maximum trapping force can be obtained by differentiating the particle--vortex binding energy with respect to their separation distance and is estimated as $F_v\simeq\rho_s\kappa^2/3\pi$~\cite{Meichle-2014-RSI}. As the ring shrinks, its speed relative to the normal fluid increases, and the resulting Stokes drag $F_d$ can pull the particle off the core once $F_d$ exceeds $F_v$. Setting $F_d\!=\!F_v$ gives a detrapping threshold velocity $v_{th}\!=\!\rho_s\kappa^2/18\pi^2 a_p\mu_n$. For a typical D$_2$ particle with $a_p\simeq1~\mu$m, $v_{th}$ is only a few mm/s~\cite{Tang-2023-NatCommun}. However, in our observed events, especially near the end time $t_f$ of the trajectories, the particle speed $v_p(t_f)$ often exceeds this value by orders of magnitude, making sustained trapping on a singly quantized core impossible. In contrast, for a multiquantum core the trapping force scales up by a factor of $n^2$, and so does $v_{th}$. In Fig.~\ref{Fig4}, we plot the ratio $v_p(t_f)/v_{th}$ for all collected events. It is clear that this ratio remains below unity for all cases, which nicely explains the observed sustained trapping. These results provide strong support for a truly multiquantum core.

A key remaining puzzle is how an apparent multiquantum vortex ring could remain intact against splitting. In other quantum-fluid platforms, theory has indicated that multiply quantized vortices can be stabilized when the core is ``filled'' or otherwise endowed with internal structure, for example by a second component in two-component BECs or by strong mass loading in highly imbalanced mixtures~\cite{Ruostekoski2004_PRA,Patrick2023_PRR,Richaud2023_PRA}. In our experiment, the only evident additional structure is a micron-sized frozen D$_2$ tracer bound to the vortex core. There is currently no framework describing how such a trapped particle could stabilize a multiquantum core. If particle loading can provide stabilization at all, the same physics might also bind an extremely compact bundle of singly quantized rings, with separations far below the D$_2$ particle size. In that limit, a true multiquantum core and a tightly packed bundle would be experimentally indistinguishable, since our smallest spatial resolution is set by $a_p\simeq1~\mu$m. However, given that the inferred ring radius is far larger than $a_p$, it is difficult to see how a single trapped particle could stabilize the core (or hold an extremely compact ring bundle) along the \emph{entire} circumference rather than merely affecting a small segment. We hope this discrepancy will motivate future research efforts to determine whether single-particle loading can modify the relevant core instability pathways~\cite{Patrick2022_PRR}, or whether an entirely different stabilization mechanism is at work.


\section*{References}
\bibliographystyle{naturemag}
\bibliography{ref_ptv_events}

\begin{thebibliography}{10}
\expandafter\ifx\csname url\endcsname\relax
  \def\url#1{\texttt{#1}}\fi
\expandafter\ifx\csname urlprefix\endcsname\relax\def\urlprefix{URL }\fi
\providecommand{\bibinfo}[2]{#2}
\providecommand{\eprint}[2][]{\url{#2}}

\bibitem{Tsubota-2013-book}
\bibinfo{author}{Tsubota, M.} \& \bibinfo{author}{Kasamatsu, K.}
\newblock \emph{\bibinfo{title}{Quantized vortices and quantum turbulence}},
  \bibinfo{pages}{283--299} (\bibinfo{publisher}{Springer Berlin Heidelberg},
  \bibinfo{address}{Berlin, Germany}, \bibinfo{year}{2013}).

\bibitem{RevModPhys.66.1125}
\bibinfo{author}{Blatter, G.}, \bibinfo{author}{Feigel'man, M.~V.},
  \bibinfo{author}{Geshkenbein, V.~B.}, \bibinfo{author}{Larkin, A.~I.} \&
  \bibinfo{author}{Vinokur, V.~M.}
\newblock \bibinfo{title}{Vortices in high-temperature superconductors}.
\newblock \emph{\bibinfo{journal}{Rev. Mod. Phys.}}
  \textbf{\bibinfo{volume}{66}}, \bibinfo{pages}{1125--1388}
  (\bibinfo{year}{1994}).

\bibitem{Greenstein-1970-Nature}
\bibinfo{author}{Greenstein, G.}
\newblock \bibinfo{title}{Superfluid turbulence in neutron stars}.
\newblock \emph{\bibinfo{journal}{Nature}} \textbf{\bibinfo{volume}{227}},
  \bibinfo{pages}{791--794} (\bibinfo{year}{1970}).

\bibitem{pines1985superfluidity}
\bibinfo{author}{Pines, D.} \& \bibinfo{author}{Alpar, M.~A.}
\newblock \bibinfo{title}{Superfluidity in neutron stars}.
\newblock \emph{\bibinfo{journal}{Nature}} \textbf{\bibinfo{volume}{316}},
  \bibinfo{pages}{27--32} (\bibinfo{year}{1985}).

\bibitem{zurek1985cosmological}
\bibinfo{author}{Zurek, W.~H.}
\newblock \bibinfo{title}{Cosmological experiments in superfluid helium?}
\newblock \emph{\bibinfo{journal}{Nature}} \textbf{\bibinfo{volume}{317}},
  \bibinfo{pages}{505--508} (\bibinfo{year}{1985}).

\bibitem{tilley2019superfluidity}
\bibinfo{author}{Tilley, D.~R.}
\newblock \emph{\bibinfo{title}{Superfluidity and superconductivity}}
  (\bibinfo{publisher}{Routledge}, \bibinfo{address}{London},
  \bibinfo{year}{2019}).

\bibitem{Landau1941PR}
\bibinfo{author}{Landau, L.}
\newblock \bibinfo{title}{{Theory of the superfluidity of helium II}}.
\newblock \emph{\bibinfo{journal}{Phys. Rev.}} \textbf{\bibinfo{volume}{60}},
  \bibinfo{pages}{356--358} (\bibinfo{year}{1941}).

\bibitem{vinen2002quantum}
\bibinfo{author}{Vinen, W.} \& \bibinfo{author}{Niemela, J.}
\newblock \bibinfo{title}{Quantum turbulence}.
\newblock \emph{\bibinfo{journal}{J. Low Temp. Phys.}}
  \textbf{\bibinfo{volume}{128}}, \bibinfo{pages}{167--231}
  (\bibinfo{year}{2002}).

\bibitem{vinen1957mutual}
\bibinfo{author}{Vinen, W.~F.}
\newblock \bibinfo{title}{Mutual friction in a heat current in liquid {helium
  II III}. {Theory of the mutual friction}}.
\newblock \emph{\bibinfo{journal}{Proc. R. Soc. A}}
  \textbf{\bibinfo{volume}{242}}, \bibinfo{pages}{493--515}
  (\bibinfo{year}{1957}).

\bibitem{Barenghi1983}
\bibinfo{author}{Barenghi, C.~F.}, \bibinfo{author}{Donnelly, R.~J.} \&
  \bibinfo{author}{Vinen, W.~F.}
\newblock \bibinfo{title}{Friction on quantized vortices in helium {II}. a
  review}.
\newblock \emph{\bibinfo{journal}{J. Low Temp. Phys.}}
  \textbf{\bibinfo{volume}{52}}, \bibinfo{pages}{189--247}
  (\bibinfo{year}{1983}).

\bibitem{Schwartz1985}
\bibinfo{author}{Schwarz, K.~W.}
\newblock \bibinfo{title}{Three-dimensional vortex dynamics in superfluid
  $^4${He}: Line-line and line-boundary interactions}.
\newblock \emph{\bibinfo{journal}{Phys. Rev. B}} \textbf{\bibinfo{volume}{31}},
  \bibinfo{pages}{5782--5804} (\bibinfo{year}{1985}).

\bibitem{Bewley-2006-Nature}
\bibinfo{author}{Bewley, G.~P.}, \bibinfo{author}{Lathrop, D.~P.} \&
  \bibinfo{author}{Sreenivasan, K.~R.}
\newblock \bibinfo{title}{Superfluid helium: Visualization of quantized
  vortices}.
\newblock \emph{\bibinfo{journal}{Nature}} \textbf{\bibinfo{volume}{441}},
  \bibinfo{pages}{588} (\bibinfo{year}{2006}).

\bibitem{Guo-2014-PNAS}
\bibinfo{author}{Guo, W.}, \bibinfo{author}{La~Mantia, M.},
  \bibinfo{author}{Lathrop, D.~P.} \& \bibinfo{author}{Van~Sciver, S.~W.}
\newblock \bibinfo{title}{Visualization of two-fluid flows of superfluid
  helium-4}.
\newblock \emph{\bibinfo{journal}{Proc. Natl Acad. Sci. USA}}
  \textbf{\bibinfo{volume}{111}}, \bibinfo{pages}{4653--4658}
  (\bibinfo{year}{2014}).

\bibitem{Mantia-PRB-2014}
\bibinfo{author}{La~Mantia, M.} \& \bibinfo{author}{Skrbek, L.}
\newblock \bibinfo{title}{Quantum turbulence visualized by particle dynamics}.
\newblock \emph{\bibinfo{journal}{Phys. Rev. B}} \textbf{\bibinfo{volume}{90}},
  \bibinfo{pages}{014519} (\bibinfo{year}{2014}).

\bibitem{Mastracci-2019-PRF}
\bibinfo{author}{Mastracci, B.} \& \bibinfo{author}{Guo, W.}
\newblock \bibinfo{title}{Characterizing vortex tangle properties in
  steady-state {He II} counterflow using particle tracking velocimetry}.
\newblock \emph{\bibinfo{journal}{Phys. Rev. Fluids}}
  \textbf{\bibinfo{volume}{4}}, \bibinfo{pages}{023301} (\bibinfo{year}{2019}).

\bibitem{Fonda-2019-PNAS}
\bibinfo{author}{Fonda, E.}, \bibinfo{author}{Sreenivasan, K.~R.} \&
  \bibinfo{author}{Lathrop, D.~P.}
\newblock \bibinfo{title}{Reconnection scaling in quantum fluids}.
\newblock \emph{\bibinfo{journal}{Proc. Natl Acad. Sci. USA}}
  \textbf{\bibinfo{volume}{116}}, \bibinfo{pages}{1924--1928}
  (\bibinfo{year}{2019}).

\bibitem{Tang-2023-NatCommun}
\bibinfo{author}{Tang, Y.} \emph{et~al.}
\newblock \bibinfo{title}{Imaging quantized vortex rings in superfluid helium
  to evaluate quantum dissipation}.
\newblock \emph{\bibinfo{journal}{Nat. Commun.}} \textbf{\bibinfo{volume}{14}},
  \bibinfo{pages}{2941} (\bibinfo{year}{2023}).

\bibitem{Peretti2023SciAdv}
\bibinfo{author}{Peretti, C.}, \bibinfo{author}{Vessaire, J.},
  \bibinfo{author}{Durozoy, E.} \& \bibinfo{author}{Gibert, M.}
\newblock \bibinfo{title}{Direct visualization of the quantum vortex lattice
  structure, oscillations, and destabilization in rotating {$^4$He}}.
\newblock \emph{\bibinfo{journal}{Sci. Adv.}} \textbf{\bibinfo{volume}{9}},
  \bibinfo{pages}{eadh2899} (\bibinfo{year}{2023}).

\bibitem{Minowa2025NatPhys}
\bibinfo{author}{Minowa, Y.} \emph{et~al.}
\newblock \bibinfo{title}{Direct excitation of {Kelvin} waves on quantized
  vortices}.
\newblock \emph{\bibinfo{journal}{Nat. Phys.}} \textbf{\bibinfo{volume}{21}},
  \bibinfo{pages}{233--238} (\bibinfo{year}{2025}).

\bibitem{Donnelly-1991-B}
\bibinfo{author}{Donnelly, R.~J.}
\newblock \emph{\bibinfo{title}{Quantized vortices in helium {II}}},
  vol.~\bibinfo{volume}{2} (\bibinfo{publisher}{Cambridge University Press},
  \bibinfo{address}{Cambridge, UK}, \bibinfo{year}{1991}).

\bibitem{Mastracci-2018-RSI}
\bibinfo{author}{Mastracci, B.} \& \bibinfo{author}{Guo, W.}
\newblock \bibinfo{title}{An apparatus for generation and quantitative
  measurement of homogeneous isotropic turbulence in {He II}}.
\newblock \emph{\bibinfo{journal}{Rev. Sci. Instrum.}}
  \textbf{\bibinfo{volume}{89}}, \bibinfo{pages}{015107}
  (\bibinfo{year}{2018}).

\bibitem{Tang-2021-PNAS}
\bibinfo{author}{Tang, Y.}, \bibinfo{author}{Bao, S.} \& \bibinfo{author}{Guo,
  W.}
\newblock \bibinfo{title}{Superdiffusion of quantized vortices uncovering
  scaling laws in quantum turbulence}.
\newblock \emph{\bibinfo{journal}{Proc. Natl Acad. Sci. USA}}
  \textbf{\bibinfo{volume}{118}}, \bibinfo{pages}{e2021957118}
  (\bibinfo{year}{2021}).

\bibitem{Sbalzarini-2005-JSB}
\bibinfo{author}{Sbalzarini, I.~F.} \& \bibinfo{author}{Koumoutsakos, P.}
\newblock \bibinfo{title}{Feature point tracking and trajectory analysis for
  video imaging in cell biology}.
\newblock \emph{\bibinfo{journal}{J. Struct. Biol.}}
  \textbf{\bibinfo{volume}{151}}, \bibinfo{pages}{182--195}
  (\bibinfo{year}{2005}).

\bibitem{Stasiak-2025-PNAS}
\bibinfo{author}{Stasiak, P.~Z.} \emph{et~al.}
\newblock \bibinfo{title}{Experimental and theoretical evidence of universality
  in superfluid vortex reconnections}.
\newblock \emph{\bibinfo{journal}{Proc. Natl Acad. Sci. USA}}
  \textbf{\bibinfo{volume}{122}}, \bibinfo{pages}{e2426064122}
  (\bibinfo{year}{2025}).

\bibitem{Kivotides-2000-Science}
\bibinfo{author}{Kivotides, D.}, \bibinfo{author}{Barenghi, C.~F.} \&
  \bibinfo{author}{Samuels, D.~C.}
\newblock \bibinfo{title}{Triple vortex ring structure in superfluid {helium
  II}}.
\newblock \emph{\bibinfo{journal}{Science}} \textbf{\bibinfo{volume}{290}},
  \bibinfo{pages}{777--779} (\bibinfo{year}{2000}).

\bibitem{Yui-2020-PRL}
\bibinfo{author}{Yui, S.}, \bibinfo{author}{Kobayashi, H.},
  \bibinfo{author}{Tsubota, M.} \& \bibinfo{author}{Guo, W.}
\newblock \bibinfo{title}{Fully coupled dynamics of the two fluids in
  superfluid {$^4$He}: Anomalous anisotropic velocity fluctuations in
  counterflow}.
\newblock \emph{\bibinfo{journal}{Phys. Rev. Lett.}}
  \textbf{\bibinfo{volume}{124}}, \bibinfo{pages}{155301}
  (\bibinfo{year}{2020}).

\bibitem{Galantucci-2020-EPJ}
\bibinfo{author}{Galantucci, L.}, \bibinfo{author}{Baggaley, A.~W.},
  \bibinfo{author}{Barenghi, C.~F.} \& \bibinfo{author}{Krstulovic, G.}
\newblock \bibinfo{title}{A new self-consistent approach of quantum turbulence
  in superfluid helium}.
\newblock \emph{\bibinfo{journal}{Eur. Phys. J. Plus}}
  \textbf{\bibinfo{volume}{135}}, \bibinfo{pages}{1--28}
  (\bibinfo{year}{2020}).

\bibitem{Landau-book}
\bibinfo{author}{Landau, L.~D.} \& \bibinfo{author}{Lifshitz, E.~M.}
\newblock \emph{\bibinfo{title}{Fluid Mechanics}}, vol.~\bibinfo{volume}{6}
  (\bibinfo{publisher}{{Pergamon Press}}, \bibinfo{address}{Oxford},
  \bibinfo{year}{1987}), \bibinfo{edition}{2} edn.

\bibitem{Guo2007_JLTP}
\bibinfo{author}{Guo, W.} \& \bibinfo{author}{Maris, H.~J.}
\newblock \bibinfo{title}{Observations of the motion of single electrons in
  liquid helium}.
\newblock \emph{\bibinfo{journal}{J. Low Temp. Phys.}}
  \textbf{\bibinfo{volume}{148}}, \bibinfo{pages}{199--206}
  (\bibinfo{year}{2007}).

\bibitem{Glaberson1968}
\bibinfo{author}{Glaberson, W.~I.}, \bibinfo{author}{Strayer, D.~M.} \&
  \bibinfo{author}{Donnelly, R.~J.}
\newblock \bibinfo{title}{Model for the core of a quantized vortex line in
  {helium II}}.
\newblock \emph{\bibinfo{journal}{Phys. Rev. Lett.}}
  \textbf{\bibinfo{volume}{21}}, \bibinfo{pages}{1740--1744}
  (\bibinfo{year}{1968}).

\bibitem{Samuels-PRL-1990}
\bibinfo{author}{Samuels, D.~C.} \& \bibinfo{author}{Donnelly, R.~J.}
\newblock \bibinfo{title}{{Dynamics of the interactions of rotons with
  quantized vortices in helium II}}.
\newblock \emph{\bibinfo{journal}{Phys. Rev. Lett.}}
  \textbf{\bibinfo{volume}{65}}, \bibinfo{pages}{187--190}
  (\bibinfo{year}{1990}).

\bibitem{galantucci-2023-PRF}
\bibinfo{author}{Galantucci, L.}, \bibinfo{author}{Krstulovic, G.} \&
  \bibinfo{author}{Barenghi, C.~F.}
\newblock \bibinfo{title}{Friction-enhanced lifetime of bundled quantum
  vortices}.
\newblock \emph{\bibinfo{journal}{Phys. Rev. Fluids}}
  \textbf{\bibinfo{volume}{8}}, \bibinfo{pages}{014702} (\bibinfo{year}{2023}).

\bibitem{Meichle-2014-RSI}
\bibinfo{author}{Meichle, D.~P.} \& \bibinfo{author}{Lathrop, D.~P.}
\newblock \bibinfo{title}{Nanoparticle dispersion in superfluid helium}.
\newblock \emph{\bibinfo{journal}{Rev. Sci. Instrum.}}
  \textbf{\bibinfo{volume}{85}}, \bibinfo{pages}{073705}
  (\bibinfo{year}{2014}).

\bibitem{Ruostekoski2004_PRA}
\bibinfo{author}{Ruostekoski, J.}
\newblock \bibinfo{title}{Stable particlelike solitons with multiply quantized
  vortex lines in {Bose-Einstein} condensates}.
\newblock \emph{\bibinfo{journal}{Phys. Rev. A}} \textbf{\bibinfo{volume}{70}},
  \bibinfo{pages}{041601(R)} (\bibinfo{year}{2004}).

\bibitem{Patrick2023_PRR}
\bibinfo{author}{Patrick, S.}, \bibinfo{author}{Gupta, A.},
  \bibinfo{author}{Gregory, R.} \& \bibinfo{author}{Barenghi, C.~F.}
\newblock \bibinfo{title}{Stability of quantized vortices in two-component
  condensates}.
\newblock \emph{\bibinfo{journal}{Phys. Rev. Res.}}
  \textbf{\bibinfo{volume}{5}}, \bibinfo{pages}{033201} (\bibinfo{year}{2023}).

\bibitem{Richaud2023_PRA}
\bibinfo{author}{Richaud, A.}, \bibinfo{author}{Lamporesi, G.},
  \bibinfo{author}{Capone, M.} \& \bibinfo{author}{Recati, A.}
\newblock \bibinfo{title}{Mass-driven vortex collisions in flat superfluids}.
\newblock \emph{\bibinfo{journal}{Phys. Rev. A}}
  \textbf{\bibinfo{volume}{107}}, \bibinfo{pages}{053317}
  (\bibinfo{year}{2023}).

\bibitem{Patrick2022_PRR}
\bibinfo{author}{Patrick, S.}, \bibinfo{author}{Geelmuyden, A.},
  \bibinfo{author}{Erne, S.}, \bibinfo{author}{Barenghi, C.~F.} \&
  \bibinfo{author}{Weinfurtner, S.}
\newblock \bibinfo{title}{Origin and evolution of the multiply quantized vortex
  instability}.
\newblock \emph{\bibinfo{journal}{Phys. Rev. Res.}}
  \textbf{\bibinfo{volume}{4}}, \bibinfo{pages}{043104} (\bibinfo{year}{2022}).

\bibitem{Xu-2008-AIP}
\bibinfo{author}{Xu, T.} \& \bibinfo{author}{Van~Sciver, S.~W.}
\newblock \bibinfo{title}{Density effect of solidified hydrogen isotope
  particles on particle image velocimetry measurements of {He II} flow}.
\newblock In \emph{\bibinfo{booktitle}{AIP Conf. Proc.}}, vol.
  \bibinfo{volume}{985}, \bibinfo{pages}{191--198}
  (\bibinfo{publisher}{American Institute of Physics},
  \bibinfo{address}{Melville, NY}, \bibinfo{year}{2008}).

\bibitem{Press1992}
\bibinfo{author}{Press, W.~H.}, \bibinfo{author}{Flannery, B.~P.},
  \bibinfo{author}{Teukolsky, S.~A.} \& \bibinfo{author}{Vetterling, W.~T.}
\newblock \emph{\bibinfo{title}{Numerical {r}ecipes in C. The {a}rt of
  {s}cientific {c}omputing}} (\bibinfo{publisher}{Cambridge University Press},
  \bibinfo{address}{Cambridge, UK}, \bibinfo{year}{1992}),
  \bibinfo{edition}{3rd} edn.

\bibitem{Adachi2010}
\bibinfo{author}{Adachi, H.}, \bibinfo{author}{Fujiyama, S.} \&
  \bibinfo{author}{Tsubota, M.}
\newblock \bibinfo{title}{Steady-state counterflow quantum turbulence:
  Simulation of vortex filaments using the full {B}iot--{S}avart law}.
\newblock \emph{\bibinfo{journal}{Phys. Rev. B}} \textbf{\bibinfo{volume}{81}},
  \bibinfo{pages}{104511} (\bibinfo{year}{2010}).

\end{thebibliography}

\section*{Methods}\label{SecVI}
\noindent\textbf{1. Particle tracking velocimetry}\\
We use solidified $D_2$ particles as tracers to visualize quantized vortices in He~II. The $D_2$ tracers are generated by mixing 1\% $D_2$ gas with 99\% He gas at room temperature using a custom-built gas-handling system (see Supplementary Fig.~1 for a schematic), and then injecting the mixture into the experimental cell. The flow rate is regulated by a needle valve, while a computer-controlled solenoid valve sets the injection duration. For each injection, the solenoid valve is opened for 30--40~s, and this sequence is repeated three times at a repetition rate of 0.02~Hz. Upon injection, the $D_2$ gas solidifies into ice particles. To promote vortex generation, in some runs we turn on a chip-resistor heater (60~$\Omega$ in He~II) by applying a step-voltage pulse of 8~V amplitude and 2~ms duration.

Particle coordinates in the image plane $(x,z)$ are identified in each frame using a feature-point tracking algorithm~\cite{Sbalzarini-2005-JSB}. The horizontal and vertical velocity components, $v_x(t)$ and $v_z(t)$, are obtained from $x(t)$ and $z(t)$ using three-point central differences or, near the end of a trajectory, two-point backward differences. The particle speed along the trajectory direction is then calculated as $v_p(t)=\sqrt{v_x(t)^2+v_z(t)^2}$. The sizes of the particles can be estimated from measurements of freely settling $D_2$ particles in quiescent He~II~\cite{Tang-2023-NatCommun}. We record images of particles during free settling and extract their settling velocities $u_p^{s}$. This settling speed is resulted from the balance between the Stokes drag, $F_d=6\pi\mu_n a_p u_p^{s}$, and the net gravitational force, $F_G=(4\pi/3)a_p^{3}(\rho_p-\rho_{\mathrm{He}})g$, where $\rho_{\mathrm{He}}=0.145$~g/cm$^{3}$ is the total He~II density and $\rho_p=0.203$~g/cm$^{3}$ is the density of solid $D_2$~\cite{Xu-2008-AIP}. Equating $F_d$ and $F_G$ yields $a_p=\sqrt{9\mu_nu_p^{s}/2(\rho_p-\rho_{\mathrm{He}})g}$. Using the measured $u_p^{s}$, we can obtain the particle-radius distribution, which peaks at $a_p\!\simeq\!1.1$~$\mu$m with a root-mean-square width of about 0.2~$\mu$m under our experimental conditions~\cite{Tang-2023-NatCommun}.\\

\noindent\textbf{2. Charge estimate for the event in Fig.~1c}\\
For the event shown in Fig.~1c, the heater surface is fixed at $V_H=+10$~V. We can compute the electric-field distribution in the experimental cell using MATLAB’s Partial Differential Equation Toolbox. At the location where the particle changes its direction of motion and begins to accelerate (see Fig.~\ref{Fig1}c), the field strength is $E\simeq 1.51\times 10^{3}$~V/m. Balancing the electrostatic force $F_E=QE$ against the Stokes drag $F_d=6\pi a_p \mu_n v_p$ yields an estimate for the required particle charge, $Q=-6\pi a_p \mu_n v_p/E\simeq -8\times 10^{2}e$, where the particle radius is taken to be $a_p\simeq 1.1$~$\mu$m. As discussed in the main text, such an abrupt acquisition of hundreds of electron charge $e$ is unrealistic.\\

\noindent\textbf{3. Kinematics of singly quantized vortex rings}\\
Within the Schwarz model~\cite{Schwartz1985}, the symmetry-axis propagation speed $v_{\parallel}(R)$ and radial shrinkage rate $dR/dt$ of a singly quantized vortex ring of radius $R$ and circulation $\kappa$ in quiescent He~II are given by~\cite{Barenghi1983,Donnelly-1991-B}:
\begin{equation}\label{eq: ring horizontal velocity}
v_{\parallel}(R)=(1-\alpha'(T))\frac{\kappa}{4\pi R}\left[\ln\!\left(\frac{8R}{a_0}\right)-\frac{1}{2}\right],
\end{equation}
\begin{equation}\label{eq: ring radial velocity}
\frac{dR}{dt}=-\alpha(T)\frac{\kappa}{4\pi R}\left[\ln\!\left(\frac{8R}{a_0}\right)-\frac{1}{2}\right],
\end{equation}
where $\alpha(T)$ and $\alpha'(T)$ are the temperature-dependant mutual-friction coefficients~\cite{Schwartz1985,Barenghi1983,Donnelly-1991-B}, and $a_0\simeq 1$~\AA\ is the vortex-core radius. The speed of a vortex segment on the ring is then $v(R)=\sqrt{v_{\parallel}(R)^2+(dR/dt)^2}$. The coefficients $\alpha(T)$ and $\alpha'(T)$ are related to the microscopic mutual-friction coefficients $D(T)$ and $D_t(T)$ by~\cite{Barenghi1983,Donnelly-1991-B}:
\begin{equation}  \label{eq: alpha}
    \alpha(T) = \frac{1}{\rho_s \kappa}\frac{a}{a^2+b^2},
\end{equation}
\begin{equation}  \label{eq: alpha prime}
    \alpha'(T) = \frac{1}{\rho_s \kappa}\frac{b}{a^2+b^2},
\end{equation}
where the parameter $a$ and $b$ are defined as:
\begin{equation} \label{eq:ab}
\begin{aligned}
a &\equiv \frac{D(T)}{D(T)^2+D_t(T)^2} + \frac{1}{E} \\
b &\equiv -\frac{D_t(T)}{D(T)^2+D_t(T)^2} + \frac{1}{\rho_s \kappa}.
\end{aligned}
\end{equation}
Here the parameter $E$ is given by~\cite{Barenghi1983,Donnelly-1991-B}:
\begin{equation} \label{eq: E}
    E \equiv \frac{-4\pi\mu_n\,[\ln(L/(2\delta))+1]}{[\ln(L/(2\delta))+1]^2+\pi^2/16},
\end{equation}
where $L=3\mu_n/(\rho_n v_G)$ is the roton mean free path with $v_G=\sqrt{2k_B T/\pi m^*}$ being the average roton group velocity. In addition, $m^*\simeq 0.16\,m_4$ is the roton effective mass, $k_B$ is the Boltzmann constant, $\rho_n$ is the normal-fluid density, and $\delta$ is the viscous penetration depth in the normal fluid. Using the tabulated values of $D(T)$ and $D_t(T)$ from Barenghi \emph{et al.}~\cite{Barenghi1983}, together with $E$ evaluated as described therein (see Appendix~A of Ref.~\cite{Barenghi1983}), one can compute $\alpha(T)$ and $\alpha'(T)$. Over the temperature range of our experiments, $\alpha(T)\sim 10^{-1}$ and $\alpha'(T)\sim 10^{-2}$. Therefore, to good accuracy, we may take:
\begin{equation}\label{eq: ring horizontal velocity approx}
v(R)\simeq v_{\parallel}(R)\simeq \frac{\kappa}{4\pi R}\left[\ln\!\left(\frac{8R}{a_0}\right)-\frac{1}{2}\right],
\end{equation}
as presented in the main text. In our data analysis, however, we still compute the full segment speed $v(R)=\sqrt{v_{\parallel}(R)^2+(dR/dt)^2}$ for comparison with the measured particle speed $v_p(t)$, although the difference relative to using $v_{\parallel}(R)$ directly is negligible.\\

\noindent\textbf{4. Initial velocity fitting and ring evolution}\\
Because the logarithmic factor in Eq.~(\ref{eq: ring radial velocity}) varies only weakly as the ring shrinks, we may treat it as approximately constant over the time window of interest~\cite{Donnelly-1991-B}. We thus define $\beta=\frac{\kappa}{4\pi}\!\left[\ln\!\left(\frac{8R}{a_0}\right)-\frac{1}{2}\right]\simeq\frac{\kappa}{4\pi}\!\left[\ln\!\left(\frac{8R(0)}{a_0}\right)-\frac{1}{2}\right]$. Eq.~\eqref{eq: ring radial velocity} then reduces to $dR/dt\!\simeq\!-\alpha(T)\beta/R$, which yields $R(t)\!\simeq\!\sqrt{R(0)^2-2\alpha(T)\beta t}$ and $v_{\parallel}(t)\simeq \beta/\sqrt{R(0)^2-2\alpha(T)\beta t}$. Guided by this functional form, we fit the first a few measured particle-speed data points to $A/\sqrt{B-t}$ and determine the initial particle speed as $v_p(0)=A/\sqrt{B}$. The uncertainty in $v_p(0)$ is taken as the residual standard error of the fit. We then infer $R(0)$ from Eq.~\eqref{eq: ring horizontal velocity approx} using $v_p(0)$. With $R(0)$ determined, we obtain $R(t)$ by numerically integrating Eq.~\eqref{eq: ring radial velocity} using a fourth-order Runge--Kutta scheme~\cite{Press1992} with a time step $\Delta t=4\times 10^{-6}$~s. The corresponding evolution of $v_{\parallel}(R(t))$ then follows from Eq.~\eqref{eq: ring horizontal velocity approx}.\\

\noindent\textbf{5. Mutual friction coefficients for multiquantum vortex cores}\\
For a multiquantum vortex ring with circulation $n\kappa$, the ring kinematics is governed by effective mutual-friction coefficients $\alpha(T,n)$ and $\alpha'(T,n)$ that depend on the circulation quantum number $n$. To obtain the appropriate $n$ dependence, we follow the roton--vortex scattering analysis of Samuels and Donnelly~\cite{Samuels-PRL-1990}, which provides a clear framework for how the microscopic mutual-friction coefficients $D$ and $D_t$ (and hence $\alpha$ and $\alpha'$) scale with circulation. They showed that $D_t\propto\kappa$, and that $D$ can be written as $D=D_1+D_2$, where $D_1\propto\kappa$ arises from roton scattering by the superfluid velocity field around the vortex core, while $D_2\propto a_0$ is associated with roton absorption and re-emission at the core. In the core model of Glaberson \emph{et al.}~\cite{Glaberson1968,Barenghi1983}, one has $a_0\propto\kappa$, because the core is reached at the radius where the circulating flow attains the Landau critical velocity. Taken together, these results imply $D(T,n)=nD(T)$ and $D_t(T,n)=nD_t(T)$. Substituting these scalings into Eqs.~\eqref{eq: alpha}--\eqref{eq:ab} then yields the multiquantum mutual-friction coefficients $\alpha(T,n)$ and $\alpha'(T,n)$:
\begin{equation}  \label{eq: alpha n}
    \alpha(T,n) = \frac{1}{\rho_s n \kappa}\frac{a_n}{a_n^2+b_n^2},
\end{equation}
\begin{equation}  \label{eq: alpha prime n}
    \alpha'(T,n) = \frac{1}{\rho_s n \kappa}\frac{b_n}{a_n^2+b_n^2},
\end{equation}
where
\begin{equation} \label{eq:ab n}
\begin{aligned}
a_n &\equiv \frac{1}{n}\frac{D(T)}{D(T)^2+D_t(T)^2} + \frac{1}{E} \\
b_n &\equiv -\frac{1}{n}\frac{D_t(T)}{D(T)^2+D_t(T)^2}
     + \frac{1}{\rho_s n\kappa} = \frac{b}{n}
\end{aligned}
\end{equation}
In the large-$n$ limit, $a_n\to 1/E$ and $b_n\to 0$, so $\alpha(T,n)$ exhibits leading-order $1/n$ scaling. This asymptotic behavior is consistent with the refined FOUCAULT model of Galantucci \emph{et al.}~\cite{Galantucci-2020-EPJ} (see Eqs.~(19)--(22) therein). $\alpha'(T,n)$ scales as $1/n^2$, therefore its effect remains negligible. The propagation speed and shrinkage rate for a multiquantum ring are then given by:
\begin{equation}  \label{eq: ring horizontal velocity n}
    v_{\parallel,n}(R) \simeq \frac{n\kappa}{4\pi R}\left[\ln\left(\frac{8R}{na_0}\right)-\frac{1}{2}\right],
\end{equation}
\begin{equation}  \label{eq: ring radial velocity n}
    \frac{dR}{dt} = -\alpha(T,n)\frac{n\kappa}{4\pi R}\left[\ln\left(\frac{8R}{na_0}\right)-\frac{1}{2}\right],
\end{equation}
With $v_p(0)$ and using Eqs.~\eqref{eq: ring horizontal velocity n} and \eqref{eq: ring radial velocity n}, we can calculate $R(t)$ and $v_{\parallel,n}(t)$ for a multiquantum ring in the same way as described above for a singly quantized ring. Similar to Eq.~\eqref{eq: ring radial velocity}, Eq.~\eqref{eq: ring radial velocity n} can also be reduced to $dR/dt \simeq -\alpha(T,n)\beta_n/R$ with $\beta_n=\frac{n\kappa}{4\pi}\left[\ln\left(\frac{8R(0)}{na_0}\right)-\frac{1}{2}\right]$, yielding $R(t)^2/R(0)^2 \simeq 1-t/\tau$ with $\tau \equiv R(0)^2/\frac{\alpha(T,n)\,n\kappa}{2\pi}\!\left[\ln\!\left(\frac{8R(0)}{na_0}\right)-\frac{1}{2}\right]$.\\

\noindent\textbf{6. Effects of the trapped D$_2$ particle}\\
For a D$2$ particle trapped on a vortex ring, the particle transmits two forces to the vortex core~\cite{Tang-2023-NatCommun}: the Stokes drag due to its relative motion with the normal fluid, $F_d \simeq 6\pi \mu_n a_p v_{\parallel,n}(R)$, and its effective weight, $F_G=(\rho_{p}-\rho_{\mathrm{He}})\frac{4\pi}{3}a_p^{3}g$. To evaluate their influence on the ring kinematics, one can compare these forces with the total mutual-friction force experienced by the ring, $F_{sn} \simeq \alpha(T,n)\,\rho_s\,n\kappa\,v_{\parallel,n}(R)\,2\pi R$. For the two events shown in Fig.~\ref{Fig1}b and Fig.~\ref{Fig1}c, evaluating at the initial radius $R(0)$ gives $F_d/F_{sn}=5.01\%$ and $3.95\%$, respectively, while $F_G/F_{sn}<0.1\%$ in both cases. The same estimate for the full dataset indicates that the trapped particle has a negligible effect on ring kinematics for all events, consistent with the conclusion in Ref.~\cite{Tang-2023-NatCommun} for rings with few trapped particles.\\

\noindent\textbf{7. Schwarz model and vortex filament simulation}\\
In the Schwarz vortex-filament model~\cite{Schwartz1985}, quantized vortices are represented as zero-thickness space curves, and the microscopic vortex-core structure is neglected. A vortex filament is parameterized by its position $\bm{s}=\bm{s}(\xi,t)$, where $\xi$ is the arc length. In a pure superfluid at $T=0$, the only hydrodynamic force on a filament element is the Magnus force. Per unit length, it is
\begin{equation}\label{eq:Magnus}
\bm{f}_{\rm M}=\rho_s\kappa\,\bm{s}'\times(\bm{u}_{\rm L}-\bm{u}_{\rm s}),
\end{equation}
where $\bm{s}'=\partial \bm{s}/\partial \xi$ is the unit tangent vector, $\bm{u}_{\rm L}=\partial \bm{s}/\partial t$ is the filament velocity, and $\bm{u}_{\rm s}$ is the local superfluid velocity at the filament location. The superfluid velocity can be decomposed as: $\bm{u}_{\rm s}=\bm{u}_{\rm s0}+\bm{u}_{\rm in}$, where $\bm{u}_{\rm s0}$ accounts for the background flow and $\bm{u}_{\rm in}$ is the velocity induced by the entire vortex configuration, which can be obtained from the Biot--Savart integral over the full set of filaments $\mathcal{L}$~\cite{Adachi2010}:
\begin{equation}\label{eq:BiotSavart}
\bm{u}_{\rm in}(\bm{s},t)=\frac{\kappa}{4\pi}\int_{\mathcal{L}}
\frac{\left[\bm{s}(\xi_1,t)-\bm{s}(\xi,t)\right]\times \bm{s}'(\xi_1,t)\,d\xi_1}
{\left|\bm{s}(\xi_1,t)-\bm{s}(\xi,t)\right|^3}.
\end{equation}
Neglecting vortex-core inertia, the net force must vanish, i.e., $\bm{f}_{\rm M}=0$, and hence the vortex equation of motion is simply $\bm{u}_{\rm L}(\xi,t)=\bm{u}_{\rm s}(\bm{s},t)=\bm{u}_{\rm s0}+\bm{u}_{\rm in}(\bm{s},t)$.

At finite temperature, interactions with the normal-fluid component give rise to an additional mutual-friction force per unit length on the vortex filament~\cite{Donnelly-1991-B,Barenghi1983,Schwartz1985}:
\begin{equation}\label{eq:mf_force_gamma}
\bm{f}_{\rm sn}=
-\gamma_0\,\bm{s}'\times\big[\bm{s}'\times(\bm{u}_{\rm n}-\bm{u}_{\rm L})\big]
+\gamma_0'\,\bm{s}'\times(\bm{u}_{\rm n}-\bm{u}_{\rm L}),
\end{equation}
where $\gamma_0(T)$ and $\gamma_0'(T)$ are temperature-dependent coefficients that are algebraically related to the mutual-friction coefficients $\alpha(T)$ and $\alpha'(T)$~\cite{Donnelly-1991-B}. Neglecting core inertia as before, the force balance $\bm{f}_{\rm M}+\bm{f}_{\rm sn}=0$ then yields the standard Schwarz equation of motion~\cite{Schwartz1985}:
\begin{equation}\label{eq:EOM_finiteT}
\bm{u}_{\rm L}
=\bm{u}_{\rm s}
+\alpha\,\bm{s}'\times(\bm{u}_{\rm n}-\bm{u}_{\rm s})
-\alpha'\,\bm{s}'\times\big[\bm{s}'\times(\bm{u}_{\rm n}-\bm{u}_{\rm s})\big].
\end{equation}
For vortices in quiescent He~II, we take $\bm{u}_{\rm n}=\bm{0}$ and $\bm{u}_{\rm s0}=\bm{0}$, so that $\bm{u}_{\rm s}=\bm{u}_{\rm in}$. Equation~\eqref{eq:EOM_finiteT} then reduces to
\begin{equation}\label{eq:EOM_quiescent}
\bm{u}_{\rm L}
=\bm{u}_{\rm in}
-\alpha\,\bm{s}'\times\bm{u}_{\rm in}
+\alpha'\,\bm{s}'\times\big(\bm{s}'\times\bm{u}_{\rm in}\big).
\end{equation}

For an isolated circular vortex ring, evaluating $\bm{u}_{\rm in}$ with the usual core cutoff recovers the analytic propagation speed and shrinkage rate given in Eqs.~(\ref{eq: ring horizontal velocity}) and~(\ref{eq: ring radial velocity}). For a vortex-ring bundle, $\bm{u}_{\rm in}$ must be computed from Eq.~\eqref{eq:BiotSavart} over \emph{all} filaments in the configuration, so that each vortex segment moves under the velocity induced by the entire bundle. To solve Eq.~\eqref{eq:EOM_quiescent} numerically, we discretize each filament into Lagrangian points and adaptively maintain the point spacing within $\Delta \xi_{\min}\le \Delta \xi \le \Delta \xi_{\max}$. At each time step, $\bm{u}_{\rm in}$ is evaluated by computing the full Biot--Savart integral~\eqref{eq:BiotSavart} over the entire discretized configuration~\cite{Adachi2010}. The resulting ordinary differential equations for the point positions are advanced using a fourth-order Runge--Kutta method~\cite{Press1992} with time step $\Delta t$. In the simulations reported here, we use $\Delta t=1.44\times 10^{-5}$~s, $\Delta \xi_{\max}=2.25\times 10^{-6}$~m, and $\Delta \xi_{\min}=1.25\times 10^{-6}$~m.

\section*{Data Availability}
\noindent The authors declare that the data supporting the findings of this study are available within the paper and its supplementary information files.

\section*{Code availability}
\noindent All computer codes used in this study are available from the corresponding author upon reasonable request.

\section*{Acknowledgments}
The authors would like to thank Luca Galantucci and Giorgio Krstulovic for valuable discussions and for showing their preliminary simulation results on few-ring bundles. The authors acknowledge the support from the Gordon and Betty Moore Foundation through Grant DOI 10.37807/gbmf11567 and the U.S. Department of Energy under Grant No. DE-SC0020113. The work was conducted at the National High Magnetic Field Laboratory at Florida State University, which is supported by the National Science Foundation Cooperative Agreement No. DMR-2128556 and the state of Florida.

\section*{Author contributions}
\noindent W.G. designed and supervised the research and wrote the paper; Y.X. and Y.A. conducted the experiment; S.I. performed the numerical simulations; All authors participated in the result analysis and paper revision.

\section*{Competing interests}
\noindent The authors declare no competing interests.

\end{document}